\documentclass[twocolumn,fleqn]{article}

\usepackage{geometry}
\usepackage{doi}

\usepackage{titlesec}
\titleformat*{\section}{\large\bfseries}
\titleformat*{\subsection}{\normalsize\bf}

\usepackage{graphicx}

\usepackage{hyperref}
\hypersetup{colorlinks=true,linkcolor=blue,citecolor=blue}
\usepackage{cite}

\usepackage{times}
\usepackage[labelfont=bf]{caption}

\usepackage{abstract}

\begin{document}
\title{\textbf{ Enhanced discrimination of high-dimensional quantum states by concatenated optimal measurement strategies}}

\author{\normalsize \textbf{M. A. Sol\'is-Prosser$^{1,*}$,\,
O. Jim\'enez$^2$,\, A. Delgado$^{3,4}$ and L. Neves}$^{5,\dagger}$
\vspace{2mm} \\
\textit{\small{$^1$Departamento de Ciencias F\'isicas, Universidad de La Frontera, Temuco, Chile}} \\[-1mm]
\textit{\small{$^2$Centro de \'Optica e Informaci\'on Cu\'antica, Facultad de Ciencias, Universidad Mayor,}} \\[-1mm]
\textit{\small{Camino La Pir\'amide 5750, Huechuraba, Santiago, Chile}} \\[-1mm]
\textit{\small{$^3$Departamento de F\'isica, Universidad de Concepci\'on, Casilla 160-C, Concepci\'on, Chile}} \\[-1mm]
\textit{\small{$^4$Millennium Institute for Research in Optics, Universidad de Concepci\'on, 160-C Concepci\'on, Chile}} \\[-1mm]
\textit{\small{$^5$Departamento de F\'isica, Universidade Federal de Minas Gerais, Belo Horizonte, MG, Brazil}} \\[-1mm]
{\footnotesize{$^*$\textcolor{red}{ miguel.solis@ufrontera.cl} } and}  \footnotesize{$^\dagger$\textcolor{red}{ lneves@fisica.ufmg.br}}
}

\date{}


\twocolumn[
  \begin{@twocolumnfalse}
    \maketitle
\vspace{-5mm}
    \begin{abstract}

\noindent The impossibility of deterministic and error-free discrimination among nonorthogonal quantum states lies at the core of quantum theory and constitutes a primitive for secure quantum communication. Demanding determinism leads to errors, while demanding certainty leads to some inconclusiveness. One of the most fundamental strategies developed for this task is the optimal unambiguous measurement. It encompasses conclusive results, which allow for error-free state retrodictions with the maximum success probability, and inconclusive results, which are discarded for not allowing perfect identifications. Interestingly, in high-dimensional Hilbert spaces the inconclusive results may contain valuable information about the input states. Here, we theoretically describe and experimentally demonstrate the discrimination of nonorthogonal states from both conclusive and inconclusive results in the optimal unambiguous strategy, by concatenating a minimum-error measurement at its inconclusive space. Our implementation comprises 4- and 9-dimensional spatially encoded photonic states. By accessing the inconclusive space to retrieve the information that is wasted in the conventional protocol, we achieve significant increases of up to a factor of 2.07 and 3.73, respectively, in the overall probabilities of correct retrodictions. The concept of concatenated optimal measurements demonstrated here can be extended to other strategies and will enable one to explore the full potential of high-dimensional nonorthogonal states for quantum communication with larger alphabets. \\

\vspace{-3mm}
\noindent \rule{\linewidth}{.5pt}
\vspace{3mm}
    \end{abstract}
  \end{@twocolumnfalse}
]

\section{Introduction}
\label{intro}
Deterministic and error-free discrimination of quantum states from single-shot measurements is a basic requirement for quantum information processing. The impossibility of doing so if the states are not orthogonal is a fundamental trait of quantum mechanics. Remarkably, these two conflicting aspects can be used in a favorable way, for instance, to implement secure quantum communication \cite{Bennett92,Phoenix00,Gisin02,Renes04}. For the task of discriminating between nonorthogonal states, if we demand determinism, we must give up certainty, and vice versa; what remains is to optimize the process for each scenario. In the first, the minimum error (ME) measurement has been developed to always provide a state retrodiction, minimizing the probability of incorrect answers \cite{Yuen75}. In the second, the optimal unambiguous discrimination (UD) was conceived to provide correct retrodictions at the expense of obtaining inconclusive answers in a minimum fraction of trials \cite{Ivanovic87}.
 
Since its conception, the optimal UD measurement became a powerful tool for many advances in quantum information science, including UD-based protocols for quantum key distribution \cite{Bennett92}, entanglement concentration \cite{Chefles88-1}, quantum cloning \cite{Duan98}, quantum teleportation \cite{Roa03}, quantum random number generation \cite{Brask17}, and quantum algorithms \cite{Bergou03}. This strategy applies only for linearly independent states \cite{Chefles88-1} and can be understood as a probabilistic transformation encompassing two possible results: (i) a \emph{conclusive} result, which maps the input states into orthogonal ones (with the maximum average success probability), so they can be unambiguously discriminated; (ii) an \emph{inconclusive} result, mapping the inputs into less distinguishable linearly dependent states, which are not amenable to a further unambiguous discrimination attempt. 

\begin{figure*}[t]
\centerline{\includegraphics[width=.81\textwidth]{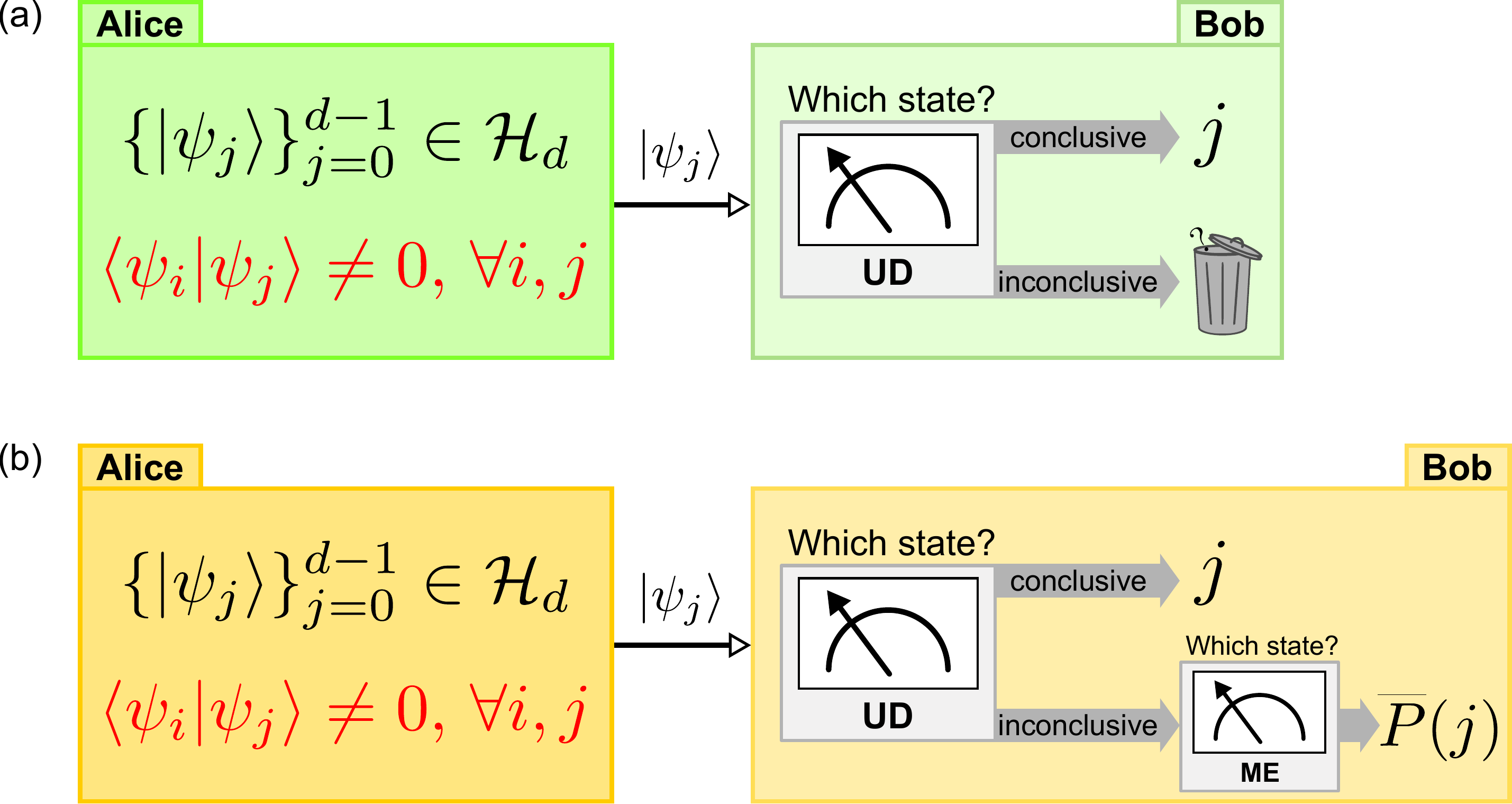}}
\caption{\label{fig:scheme}  Quantum communication schemes using $d$ nonorthogonal states as the $d$-ary ``alphabet''. Alice prepares a quantum system in a given state of the set $\{|\psi_j\rangle\}$ and sends it to Bob who tries to identify which state he received through a single-shot measurement. In (a), Bob performs a conventional UD measurement, where a conclusive result leads to a perfect identification, and an inconclusive one is discarded as useless resource. In (b), Bob performs the CUD protocol, concatenating the ME measurement at the inconclusive space. If a perfect identification fails, he may extract the remaining information from the inconclusive results with the minimum average probability of error, increasing the overall probability of correctly determining which state Alice sent him.}
\end{figure*}

Strictly speaking, the inconclusive states are of no interest for the purpose of the UD protocol, so they are, in principle, discarded as useless resources. However, in $d$-dimensional Hilbert spaces with $d>2$, unless these states are all the same, they will carry useful information about the input ones. Thus, we can still discriminate them with another suitable measurement strategy (e.g., ME) and make retrodictions that, although imperfect, would increase the information gain about the inputs. The concatenation of a strategy at the inconclusive space will enable state retrodictions from both conclusive and inconclusive results in optimal UD; such protocol will be referred here as concatenated UD (CUD), and both UD and CUD protocols are illustrated in Fig.~\ref{fig:scheme} as a quantum communication problem. The possibility of concatenating optimized measurements has been the subject of intense investigation in quantum detection theory \cite{Chefles88-1,Peres98,Chefles00,Sun01,Roa11,Jimenez11,Zhang14}, but in practice it became viable only recently with the development of experimental techniques for preparation, manipulation, and measurement of high-dimensional quantum systems. In addition to its fundamental interest, recent studies show that this concatenation provides significant improvements in probabilistic realizations of protocols like teleportation \cite{Neves12}, entanglement swapping \cite{Prosser14}, and dense coding \cite{Kogler17}; it will also have impact on high-dimensional quantum cryptography \cite{Kogler17}.

Unambiguous discrimination in high dimensions was experimentally demonstrated first by Mohseni \emph{et al.}\ for three states encoded in three longitudinal modes of an eight-port optical interferometer \cite{Mohseni04}. Recently, Becerra \emph{et al.}\ implemented UD of four nonorthogonal coherent states \cite{Becerra13} and Agnew \emph{et al.}\ performed UD between $d$ $d$-dimensional states encoded in the orbital angular momentum of single photons, for $d$ up to 14 \cite{Agnew14}. None of these experiments, however, have probed the inconclusive space to implement the concatenated protocol described above. Therefore, the potential of high-dimensional systems, which are greatly beneficial to quantum communication \cite{Fujiwara03,Martinez18} and computation \cite{Muthukrishnan00,Lanyon09}, has not been fully explored in quantum state discrimination.

In this work, we theoretically describe and experimentally demonstrate the concatenation of optimal UD with a ME measurement at the inconclusive space to discriminate nonorthogonal symmetric states, a fundamental resource for a variety of tasks in quantum information processing (e.g., see \cite{Phoenix00,Renes04,Chefles88-1,Neves12,Prosser14,Kogler17,Fujiwara03,Bavaresco18}). Our CUD protocol can be implemented for any individual high-dimensional quantum system, regardless the underlying physical platform. Here, we implement it for 4- and 9-dimensional spatially encoded photonic states \cite{Neves05}. For dozens of tests performed, we obtained---even under unavoidable experimental imperfections---higher overall probabilities of correct retrodictions in the concatenated scheme than is theoretically possible for the conventional one, achieving striking increases for this figure of merit. Our results show that the concatenation of optimized measurements is crucial to not waste useful information in high dimensions. In addition to the CUD protocol demonstrated here, we also discuss other possible concatenations between different strategies that will enhance other figures of merit in the state-discrimination process.

\section{Methods}
\subsection{Theory}   \label{subsec:Theory}

The goal of quantum state discrimination is to determine, from a single shot measurement, in which state a quantum system has been prepared, knowing the set of possible states and their {\it a priori} probabilities \cite{Chefles00}. Our implementation of CUD comprises sets of equally likely, linearly independent symmetric states, defined by \cite{Chefles88-2}
\begin{equation}   \label{eq:sym_states_in}
|\psi_j\rangle=\sum_{k=0}^{d-1}c_k\omega^{jk}|k\rangle,\;\;\;\;\; j=0,\ldots,d-1,
\end{equation}
where $|c_k|\neq 0$ $\forall k$ (with $\sum_k|c_k|^2=1$), $\omega=\exp(2\pi i/d)$, and $\{|k\rangle\}_{k=0}^{d-1}$ is an orthonormal basis in the $d$-dimensional Hilbert space, $\mathcal{H}_d$. An optimal error-free identification of a state from this set starts by applying a transformation specified by the Kraus operators  $\{\hat{A}_\perp,\hat{A}_?\}$ given by \cite{Jimenez11}
\begin{equation}   \label{eq:Kraus}
  \begin{array}{l}
\hat{A}_\perp=\displaystyle\sum_{n=0}^{d-1} \frac{c_{\rm min}}{|c_n|} \,e^{-i\arg(c_n)}|n\rangle\langle n|,\\[7mm]
\hat{A}_?=\displaystyle\sum_{n=0}^{d-1} \sqrt{1-\frac{c_{\rm min}^2}{|c_n|^2}} \,e^{-i\arg(c_n)}|n\rangle\langle n|,
  \end{array}
\end{equation}
where $\hat{A}_\perp$ and $\hat{A}_?$ denote the conclusive and inconclusive operators, respectively, and $c_{\rm min}\equiv\min\{|c_k|\}_{k=0}^{d-1}$. Their action transforms the inputs of Eq.~(\ref{eq:sym_states_in}) as 
\begin{equation}   \label{eq:maps}
  \begin{array}{l}
\hat{A}_\perp|\psi_j\rangle=\sqrt{p_\perp}\displaystyle|\phi_j^\perp\rangle,\\[5mm]
\hat{A}_?|\psi_j\rangle=\sqrt{p_?}\displaystyle|\phi_j^?\rangle,
  \end{array}
\end{equation}
 where  
\begin{equation}   \label{eq:optimal_p}
p_\perp=1-p_?=dc_{\rm min}^2
\end{equation}
is the optimal probability of a conclusive transformation, and
\begin{equation}   \label{eq:phi_states}
  \begin{array}{l}
\displaystyle|\phi_j^\perp\rangle=\frac{1}{\sqrt{d}}
\sum^{d-1}_{k=0}\omega^{jk}|k\rangle,\\[7mm]
\displaystyle|\phi_j^?\rangle=
\sum^{d-1}_{k=0}\sqrt{\frac{|c_k|^2-c_{\rm min}^2}{p_?}}\omega^{jk}|k\rangle.
  \end{array}
\end{equation}
The states $\{|\phi_j^\perp\rangle\}$ are orthogonal whereas $\{|\phi_j^?\rangle\}$ are linearly dependent. In the conclusive case, a perfect retrodiction of $|\psi_j\rangle$ is achieved with the projective ME measurement 
\begin{equation}    \label{eq:Fourier}
\hat{\Pi}^{\rm\textsc{me}}_{n}=\hat{\mathcal{F}}|n\rangle\langle n|\hat{\mathcal{F}}^{-1},\;\;\;\;\; n=0,\ldots,d-1,
\end{equation}
 where $\hat{\mathcal{F}}=\sum_{j,k=0}^{d-1}\omega^{jk}|j\rangle\langle k|/\sqrt{d}$ is the quantum Fourier transform on $\mathcal{H}_d$, so that 
\begin{equation}
\langle\phi_j^\perp|\hat{\Pi}^{\rm\textsc{me}}_{n}|\phi_j^\perp\rangle=\delta_{jn},
\end{equation}
where $\delta_{jn}$ denotes the Kronecker delta. Otherwise, for an inconclusive transformation, the output states cannot be perfectly identified anymore.
In the conventional UD protocol (Fig.~\hyperref[fig:scheme]{\ref{fig:scheme}(a)}), there is no concern about these states, so one can retrodict nothing but a random guess from an inconclusive event. On the other hand, in the CUD protocol (Fig.~\hyperref[fig:scheme]{\ref{fig:scheme}(b)}) one accesses the inconclusive space, $\mathcal{H}_?$, by applying a strategy for discriminating linearly dependent states which, although imperfect, may provide better retrodictions than a random guess. Here, we implement the ME measurement ($\hat{\Pi}_?^{\rm\textsc{me}}$). The inconclusive states $\{|\phi_j^?\rangle\}$ defined in Eq.~(\ref{eq:phi_states}) form a set of $d$ equally likely symmetric states which are linearly dependent, since $\dim\mathcal{H}_?=d-\mu(c_{\rm min})$, where $\mu(c_{\rm min})$ denotes the multiplicity of $c_{\rm min}$. In this case, $\hat{\Pi}_?^{\rm\textsc{me}}$ is also given by Eq.~(\ref{eq:Fourier}): this is a projective measurement on $\mathcal{H}_d$ that, from Neumark's theorem, realizes the positive operator valued measure for ME discrimination on $\mathcal{H}_?$  \cite{Ban97,Prosser17}.

To compare the performances of the CUD and the conventional UD protocols, we must account for a figure of merit that encompasses both conclusive and inconclusive results. A straightforward choice is the overall probability of correct retrodictions, $\mathcal{P}^{\rm\textsc{(c)ud}}$, given, in each case,  by
\begin{equation}   \label{eq:Pcorr}
  \begin{array}{l}
\displaystyle\mathcal{P}^{\rm\textsc{ud}}=p_\perp P_{\perp}^{\rm \textsc{me}} + p_?\frac{1}{d},\\[7mm]
\displaystyle\mathcal{P}^{\rm\textsc{cud}}=p_\perp P_{\perp}^{\rm \textsc{me}}+p_? P_{?}^{\rm \textsc{me}},
  \end{array}
\end{equation}
where $p_\perp=1-p_?$ is given by Eq.~(\ref{eq:optimal_p}), $1/d$ corresponds to the probability from a random guess, and
\begin{equation}
P_{\ell}^{\rm \textsc{me}}= \frac{1}{d}\sum_{j=0}^{d-1}\langle\phi_j^\ell|\hat{\Pi}_{j}^{\rm\textsc{me}}|\phi_j^\ell\rangle 
\end{equation}
is the probability of correct retrodictions given by the ME measurement in $\mathcal{H}_d$ ($\ell=\perp$) or $\mathcal{H}_?$ ($\ell=?$). Using Eqs.~(\ref{eq:phi_states}) and (\ref{eq:Fourier}) it is easy to check that $P_{\perp}^{\rm \textsc{me}}=1$ (as expected, theoretically) and $P_{?}^{\rm \textsc{me}}=[\sum_k(|c_k|^2-c^2_{\rm min})^{1/2}]^2/dp_?$ $\in[1/d,1)$. If $\mu(c_{\rm min})=d-1$, then $\dim\mathcal{H}_?=1$ and  $P_{?}^{\rm \textsc{me}}=1/d$; otherwise, if $\mu(c_{\rm min})<d-1$, then $\dim\mathcal{H}_?>1$ and  $P_{?}^{\rm \textsc{me}}>1/d$. Therefore,
\begin{equation}   \label{eq:Advantage}
\mathcal{P}^{\rm\textsc{cud}}\geq\mathcal{P}^{\rm\textsc{ud}},
\end{equation}
which establishes the advantage of performing the CUD protocol whenever the dimension of the inconclusive space is larger than one.

\subsection{Experiment}

\begin{figure*}[t]
\centerline{\includegraphics[width=1\textwidth]{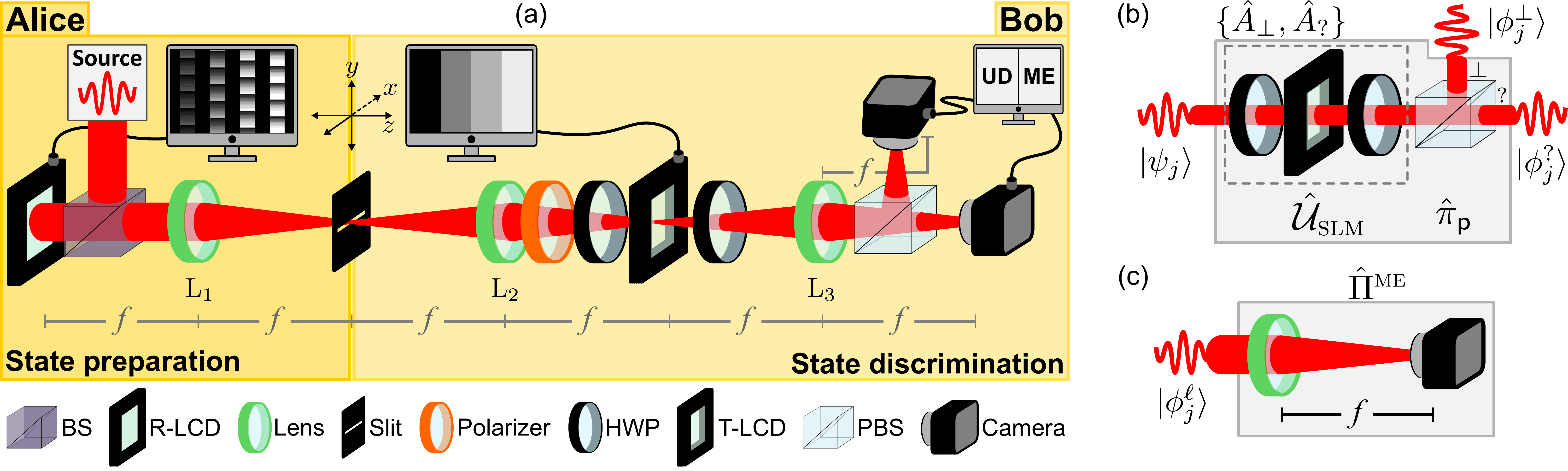}}
\caption{\label{fig:setup} (a) Experimental setup (see text for details). Lenses ${\rm L}_1$--${\rm L}_3$ have focal length $f=30\!$~cm. R(T)-LCD: reflective (transmissive) liquid crystal display; (P)BS: (polarizing) beam splitter; HWP: half-wave plate. (b) Arrangement that implements the Kraus operators of Eq.~(\ref{eq:Kraus}) which transform the input states according to Eqs.~(\ref{eq:maps})--(\ref{eq:phi_states}): $\hat{\mathcal{U}}_{\rm\textsc{slm}}$ is given by Eq.~(\ref{eq:unitary}) and $\hat{\pi}_{\rm\textsf{p}}$ is the polarization projection by the PBS. (c) Arrangement for the ME measurement given by Eq.~(\ref{eq:Fourier}).} 
\end{figure*}

Like many optical tests of quantum state discrimination \cite{Mohseni04,Prosser17,Huttner96,Barnett97,Clarke01-1,Clarke01-2,Mizuno01,Mosley06} and other quantum phenomena in general \cite{Martinez18,Mir07,Malik14,Kewming20,Sahoo20}, here we used a laser light source to demonstrate the CUD protocol outlined above. Our experimental setup is sketched in Fig.~\hyperref[fig:setup]{\ref{fig:setup}(a)}. 

In the state preparation stage (Alice), the source  consists of an expanded, collimated, and vertically polarized gaussian beam produced from a spatially filtered single-mode diode laser operating at 687$\!$~nm (e.g., see \cite{Prosser17,Prosser13}). This beam is normally incident on a reflective liquid crystal display (LCD, Holoeye PLUTO) working as a phase-only spatial light modulator (SLM), which is addressed with a computer-generated mask given by an array of $d$ blazed diffraction gratings. A typical mask is shown in Alice's computer for $d=4$; the gratings have width and separation of 18 pixels, and period of 12 pixels, for a pixel size of $8\!$~$\mu$m. The modulated beam is transmitted through the spherical lens ${\rm L}_1$ and in its focal plane the first diffraction order is selected by a slit diaphragm. The emerging light is a coherent superposition of $d$ non-overlapping spatial modes, $\{|k\rangle\}_{k=0}^{d-1}$, generated by the gratings in the $x$ direction. Each mode is modulated by a complex coefficient whose magnitude is a function of the phase depth of the grating whereas the phase is defined by its lateral displacement. With this procedure, fully described in \cite{Prosser13,Varga14}, we generate superpositions that are identical to a stream of single photons prepared in a symmetric state (Eq.~(\ref{eq:sym_states_in})).

In the state discrimination stage (Bob), the light modulated in Alice's SLM is imaged onto a transmissive LCD (Holoeye LC 2012) by a $4f$ system (lenses ${\rm L}_1$ and ${\rm L}_2$). The LCD is sandwiched by two half-wave plates (HWPs) and this set forms Bob's SLM which will modulate both phase and polarization of the incoming light as a function of the voltage (or gray level) applied to the LCD \cite{Moreno03}. At the entrance of the SLM, a polarizer is used to ensure a pure vertical polarization for this ancillary degree of freedom, so the input state will be written as $|\psi_j\rangle|{\rm\textsf{v}}\rangle$. The LCD is addressed with a mask composed by $d$ adjacent rectangular regions 8 pixels wide (for a pixel size of $36\!$~$\mu$m), where each region has a constant gray level. A typical mask is shown in Bob's computer for the $d=4$ state prepared by Alice. Since each spatial mode $|n\rangle|{\rm\textsf{v}}\rangle$ is imaged onto a single region of this mask, the global SLM operation on $|\psi_j\rangle|{\rm\textsf{v}}\rangle$ will be  $\hat{\mathcal{U}}_{\rm\textsc{slm}}=\sum_{n=0}^{d-1}|n\rangle\langle n|\otimes\hat{J}_n$, where $\hat{J}_n$ is the unitary Jones matrix of the SLM acting on the $n$-th mode.  After calibrating the LCD, we programmed $\hat{\mathcal{U}}_{\rm\textsc{slm}}$ to operate---with a good approximation---as (see \ref{appA} for technical details) 
\begin{equation}   \label{eq:unitary}
\hat{\mathcal{U}}_{\rm\textsc{slm}}=\sum_{n=0}^{d-1}|n\rangle\langle n|\otimes e^{i\varphi_n}\left[
\begin{array}{cc}
a_n & b_n \\[1mm]
-b_n & a_n
\end{array}
\right],
\end{equation}
where the $2\times 2$ matrix is written in the polarization basis $\{|{\rm\textsf{h}}\rangle=(1,0)^{\rm\textsf{T}},|{\rm\textsf{v}}\rangle=(0,1)^{\rm\textsf{T}}\}$. 
The parameter $\varphi_n$ is a phase shift and $a_n=\sqrt{1-b_n^2}$ is a polarization amplitude; both are real and depend on the gray level (\textsf{gl}) addressed to the $n$-th region of the LCD mask. Now, it is easy to check that by preparing the input states given by Eq.~(\ref{eq:sym_states_in}) with $\arg(c_n)=-\varphi_n(\textsf{gl})$ and setting $a_n(\textsf{gl})=c_{\rm min}/|c_n|$, we obtain
\begin{equation}    \label{eq:Transform}
\hat{\mathcal{U}}_{\rm\textsc{slm}}|\psi_j\rangle|{\rm\textsf{v}}\rangle
=\sqrt{p_\perp}|\phi_j^\perp\rangle|{\rm\textsf{v}}\rangle
+\sqrt{p_?}|\phi_j^?\rangle|{\rm\textsf{h}}\rangle,
\end{equation}
where $p_\ell$ and $|\phi_j^\ell\rangle$  ($\ell=\perp,?$) are defined in Eqs.~(\ref{eq:optimal_p}) and (\ref{eq:phi_states}), respectively. 
Therefore, the transformation by the Kraus operators on the inputs (see Eqs.~(\ref{eq:Kraus}) and (\ref{eq:maps})) is implemented by applying $\hat{\mathcal{U}}_{\rm\textsc{slm}}$ and performing a polarization projection, $\hat{\pi}_{\rm\textsf{p}}$, in the basis $\{|{\rm\textsf{h}}\rangle,|{\rm\textsf{v}}\rangle\}$ with a polarizing beam splitter (PBS), so that
\begin{equation}   \label{eq:post_PBS}
  \begin{array}{l}
\langle{\rm\textsf{v}}|\hat{\mathcal{U}}_{\rm\textsc{slm}}|\psi_j\rangle|{\rm\textsf{v}}\rangle
=\sqrt{p_\perp}|\phi_j^\perp\rangle\equiv\hat{A}_\perp|\psi_j\rangle,\\[5mm]
\langle{\rm\textsf{h}}|\hat{\mathcal{U}}_{\rm\textsc{slm}}|\psi_j\rangle|{\rm\textsf{v}}\rangle
=\sqrt{p_?}|\phi_j^?\rangle\equiv\hat{A}_?|\psi_j\rangle;  
\end{array}
\end{equation}
the output ``\textsf{v}'' (``\textsf{h}'') at the PBS indicates a conclusive (inconclusive) mapping with the optimal probability $p_\perp$ ($p_?$). This process is sketched in Fig.~\hyperref[fig:setup]{\ref{fig:setup}(b)}. 

To conclude the CUD protocol, a CMOS camera (Thorlabs DCC1545M) is placed at the focal plane of the lens ${\rm L}_3$ at each output of the PBS (see Fig.~\hyperref[fig:setup]{\ref{fig:setup}(a)}). We select $d$ pixels of each camera distributed along the transverse positions 
\begin{equation}   \label{eq:pixel_pos}
x_j=-\frac{\lambda fq_j}{sd},\;\;\;\;\;\; j=0,\ldots,d-1,
\end{equation} 
where $\lambda$ is the light wavelength, $f$ the lens focal length, $s$ the separation between adjacent spatial modes, and $q_j=j$ if $j\leq d/2$ or $q_j=j-d$ if $j>d/2$. As we showed in \cite{Prosser17}, this arrangement, sketched in Fig.~\hyperref[fig:setup]{\ref{fig:setup}(c)}, will implement the ME measurement given by Eq.~(\ref{eq:Fourier}) at both conclusive and inconclusive arms, since each pixel postselects a superposition of spatial modes given by $|\eta(x_j)\rangle=\frac{1}{\sqrt{d}}\sum_{k=0}^{d-1}\omega^{jk}|k\rangle=\hat{\mathcal{F}}|j\rangle$.

\section{Results}  \label{sec:Results}

\begin{figure}[t]
\centerline{\includegraphics[width=1\columnwidth]{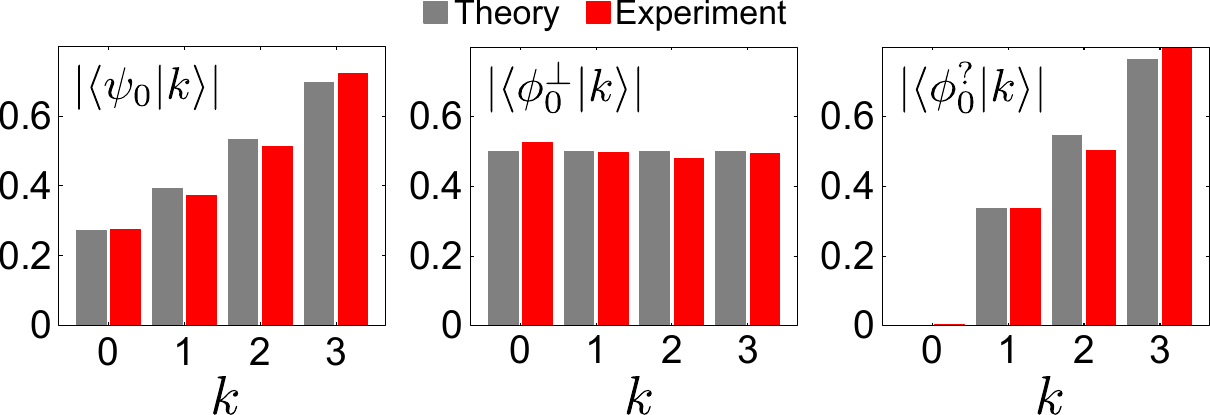}}
\caption{\label{fig:coef} Coefficient magnitudes $|\langle\bullet|k\rangle|$ as a function of the index $k=0,1,2,3$ for the states prepared ($|\psi_0\rangle$) and transformed ($|\phi_0^\perp\rangle$, $|\phi_0^?\rangle$) with the LCD masks shown in Alice's and Bob's computer screens of Fig.~\hyperref[fig:setup]{\ref{fig:setup}(a)}, respectively. The theoretical results (gray bars) are obtained from Eqs.~(\ref{eq:sym_states_in}) and (\ref{eq:phi_states}).}
\end{figure}

Let us first illustrate the effect of the Kraus operation on the input states by the arrangement shown in Fig.~\hyperref[fig:setup]{\ref{fig:setup}(b)}. For this purpose, we measured the magnitudes of the coefficients of $|\psi_0\rangle$ and $|\phi_0^\ell\rangle$ ($\ell=\perp,?$) prepared with Alice's mask and transformed with Bob's mask, respectively, shown in Fig.~\hyperref[fig:setup]{\ref{fig:setup}(a)}. This was done with a camera at the focal plane of ${\rm L}_2$ (using a not shown beam splitter) and by replacing ${\rm L}_3$ by a lens with focal length $f/2$. The measured light intensities at each camera were normalized and the results are shown in Fig.~\ref{fig:coef} along with the theoretical predictions. Up to small deviations, one observes the expected behavior for the coefficient magnitudes when the Kraus operators act on a symmetric state (see Eqs.~(\ref{eq:sym_states_in})--(\ref{eq:phi_states})). 

We performed the experiment for dimensions $d=4$ and 9. The coefficients $\{c_n\}_{n=0}^{d-1}$ of the input states (Eq.~(\ref{eq:sym_states_in})) prepared in Alice's SLM were generated with $\arg(c_n)=-\varphi_n(\textsf{gl})$, so that the phases imprinted by Bob's SLM (see Eq.~(\ref{eq:unitary})) would be canceled in order to accomplish the transformation given by Eq.~(\ref{eq:Transform}). For the magnitudes $|c_n|$ we used the two-parameter representation introduced in \cite{Prosser17}: $|c_n|^2(j_0,\xi)\propto 1-\sum_{l=j_0}^{d-1}\delta_{nl}\sqrt[d]{\xi(n-j_0+1)/(d-j_0)}$, where $j_0=1,\ldots,d-1$ and $\xi\in[0,1)$; $\xi_{\rm max}=\max\{\xi\}$ is constrained by the minimum value for $|c_n|$ that Bob's SLM can handle (see \ref{appA}). Given a coefficient index $n$ and setting $(j_0,\xi)$, one determines $|c_n|$ and encode this information in the $n$-th diffraction grating at Alice's SLM. This parametrization provides  diverse and well distinguishable sets of symmetric states. For each $d$ we tested $(d-1)\times 10$ sets.

To carry out the measurements, we first define the set of input states and the corresponding Kraus operators by specifying $d$ and the coefficient amplitudes $\{c_n\}$ according to the parametrization above. From this definition, we generate the set of masks to be addressed to the LCDs (see Fig.~\hyperref[fig:setup]{\ref{fig:setup}(a)}): $d$ masks for preparing the states and one mask to implement the unitary of Eq.~(\ref{eq:unitary}). Thereafter, we prepare and discriminate one state at a time by measuring the light intensities with the cameras at both conclusive and inconclusive arms. For each camera, the intensities are recorded at the $d$ transverse positions $\{x_i\}$ defined in Eq.~(\ref{eq:pixel_pos}); as a postprocessing, we subtract the background noise and apply a small compensation for the detection efficiency due to diffraction \cite{Prosser17}. Let $I^\ell_{ij}$ denote the resulting intensity at the position $x_i$ of the camera $\ell$ ($\ell=\perp,?$) when the input state is $|\psi_j\rangle$. Accordingly, given $|\psi_j\rangle$, $P_\ell(x_i|\psi_j)\equiv P^\ell_{ij}=I^\ell_{ij}/\sum_{k=0}^{d-1}I^\ell_{kj}$ will be the experimental conditional probabilities of correct ($i=j$) or incorrect ($i\neq j$) retrodictions at camera $\ell$, and $p(\bot|\psi_j)\equiv p_{\perp j}=\sum_{i=0}^{d-1}I^\perp_{ij}/\sum_{i=0}^{d-1}(I^\perp_{ij}+I^?_{ij})$, the probability of a conclusive result ($p_{?j}=1-p_{\perp j}$). From these probabilities we can compute all quantities of interest, in particular, those  in Eq.~(\ref{eq:Pcorr}): $[P^{\rm\textsc{me}}_{\ell}]_{\rm expt}=\sum_{j=0}^{d-1}P_{jj}^\ell/d$ and $[p_\perp]_{\rm expt}=\sum_{j=0}^{d-1}p_{\perp j}/d$. 

\begin{figure}[t]
\centerline{\includegraphics[width=1\columnwidth]{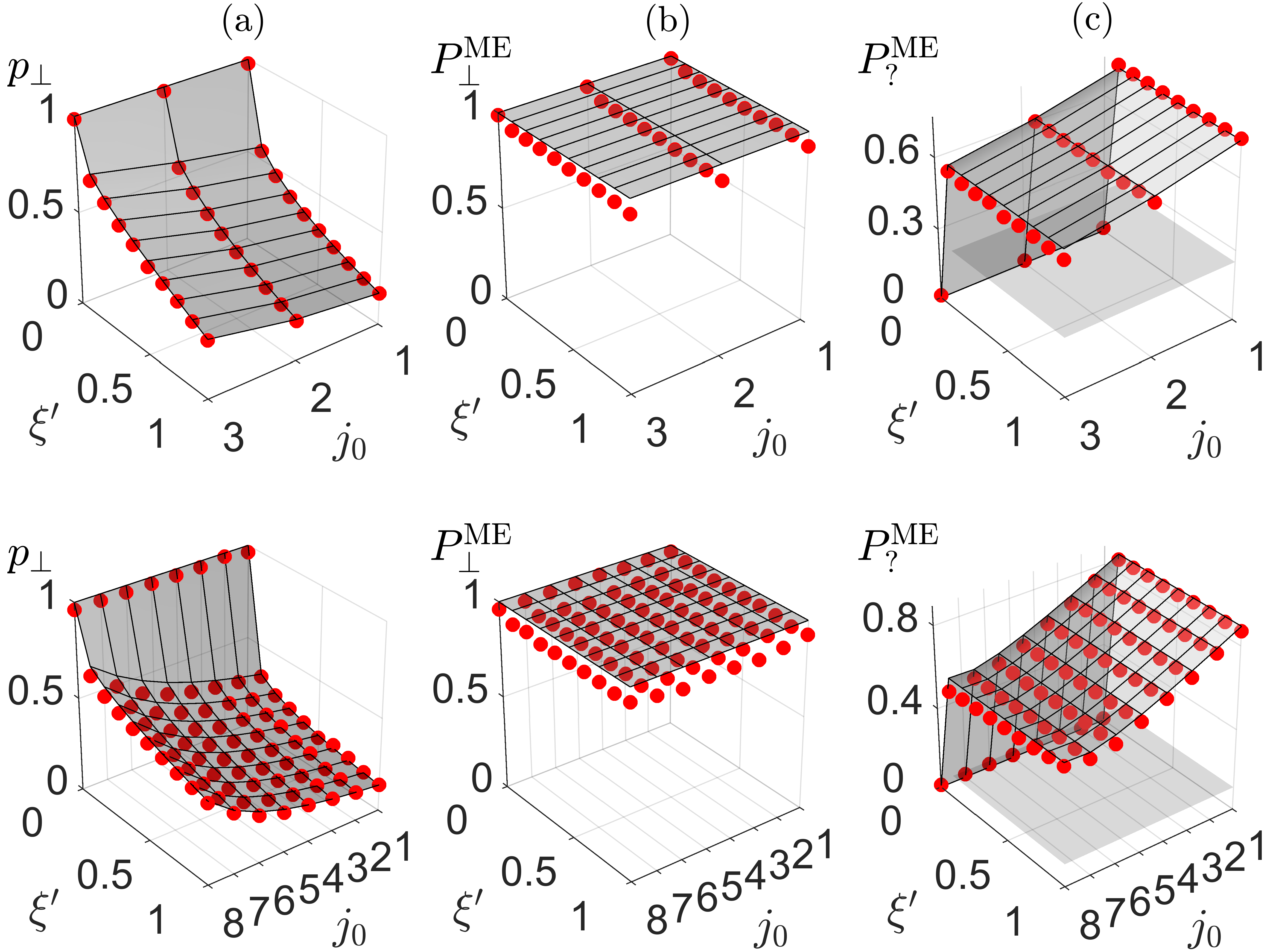}}
\caption{\label{fig:d4} Probabilities of conclusive results (a) and correct retrodictions from a conclusive (b) or inconclusive event (c) as functions of the parameters $(j_0,\xi'=\xi/\xi_{\rm max})$ that define a set of states (see text).  Top: $d=4$; bottom: $d=9$. Experiment (red points); theory (surfaces). The flat surfaces in (c) show the probability from a random guess. }
\end{figure}

Figure~\ref{fig:d4} shows how each probability involved in the protocol behaved. We plot the probabilities of conclusive results (Fig.~\hyperref[fig:d4]{\ref{fig:d4}(a)}) and correct retrodictions from a conclusive (Fig.~\hyperref[fig:d4]{\ref{fig:d4}(b)}) or inconclusive (Fig.~\hyperref[fig:d4]{\ref{fig:d4}(c)}) event, as functions of the parameters $(j_0,\xi)$ that define each tested set of states. The experimental results (red points) are in good agreement with the theoretical predictions (surfaces). The flat surfaces in Fig.~\hyperref[fig:d4]{\ref{fig:d4}(c)} represent the probability of correct state retrodictions from a random guess. By comparing them with $P^{\textsc{me}}_{?}$, one clearly observes the benefits of not discarding the inconclusive results when they occur ($\xi\neq 0$, in this case). 

For each of these probabilities, the percentage errors between the predicted and estimated ones got, in the worst cases, below 8\% for $d=4$ and 13\% for $d=9$ (for a complete analysis, see \ref{appB}). These errors are mainly caused by some degree of depolarization generated by Bob's LCD as a function of the addressed gray level \cite{Marquez:08}. This depolarization changes the transmitted/reflected intensities at the PBS, thus affecting $p_\perp$. It also introduces some decoherence in the polarization-spatial coupling, which affects the interference between the spatial modes (crucial for the ME measurement in each arm) and, consequently, the probabilities $P^{\rm\textsc{me}}_{\ell}$. Nevertheless, these  errors have not prevented the satisfactory operation of the protocol. For instance, the probabilities of conclusive results got close to the optimal ones (see Fig.~\hyperref[fig:d4]{\ref{fig:d4}(a)}); the averages of $[P^{\rm\textsc{me}}_{\perp}]_{\rm expt}$ were 0.945 and 0.921 for $d=4$ and 9, respectively;  the errors of $[P^{\rm\textsc{me}}_{?}]_{\rm expt}$ were much smaller than would be required to take these probabilities close to a random guess (see Fig.~\hyperref[fig:d4]{\ref{fig:d4}(c)}).

\begin{figure}[t]
\centerline{\includegraphics[width=1\columnwidth]{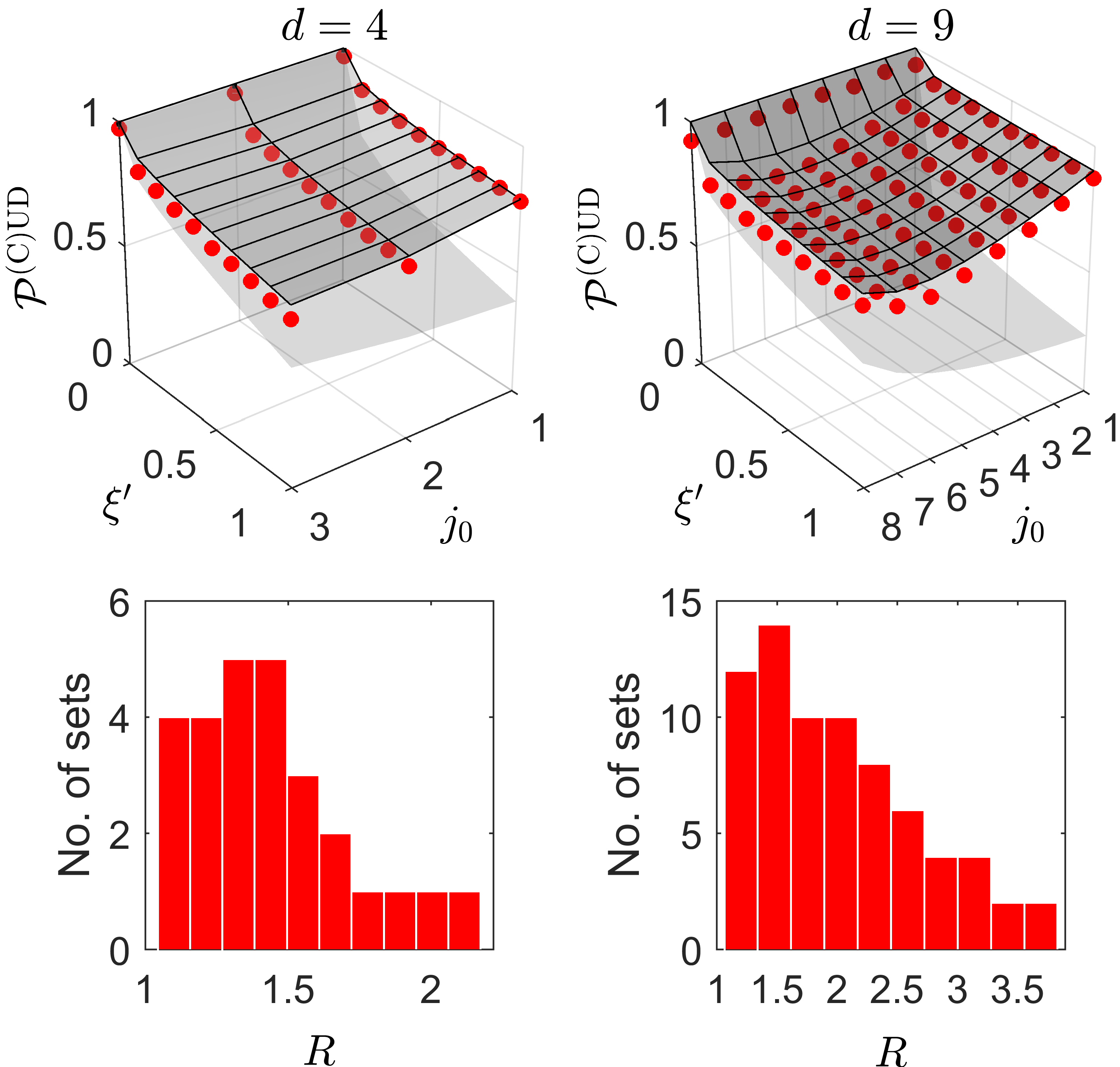}}
\caption{\label{fig:grafs} Top: overall probabilities of correct retrodictions; $\xi'=\xi/\xi_{\rm max}$. CUD: experiment (red points); theory (edged surfaces). UD: theory (non-edged surfaces). Bottom: histograms with the distribution of the enhancement ratio $R$ given by Eq.~(\ref{eq:R}).} 
\end{figure}
 
The main result of this work is shown in Fig.~\ref{fig:grafs}. In the first row, using Eq.~(\ref{eq:Pcorr}), we plot the experimentally estimated overall probabilities of correct retrodictions (red points) as a function of $(j_0,\xi)$, together with the optimal theoretical predictions for $\mathcal{P}^{\textsc{cud}}$ (edged surfaces) and $\mathcal{P}^{\textsc{ud}}$ (non-edged surfaces). Again, we observe a good agreement between theory and experiment. To quantify the contrast between our experimental results and the theoretical predictions for conventional UD, we compute the enhancement ratio 
\begin{equation}    \label{eq:R}
R=\frac{[\mathcal{P}^{\textsc{cud}}]_{\rm expt}}{ [\mathcal{P}^{\textsc{ud}}]_{\rm theor}}
\end{equation}
 for all tested sets of nonorthogonal states ($\xi\neq 0$). In the second row of Fig.~\ref{fig:grafs} we plot the corresponding histograms of $R$ for each dimension. It can be seen that $R>1$ in all cases, showing that even with the experimental errors discussed above, the probabilities obtained in our implementation were higher than is theoretically possible for conventional UD. This includes the most difficult cases of sets of states with small overlaps for which $\mathcal{P}^{\textsc{cud}}$ approximates $\mathcal{P}^{\textsc{ud}}$. The maximum values of $R$ were 2.07 for $d=4$ and 3.73 for $d=9$, indicating that remarkable increases in the overall probabilities of correct retrodictions may be achieved by concatenating the measurement strategies.

\section{Discussion and conclusion}

Our work show that there is no reason to discard the inconclusive results in the high-dimensional optimal UD measurement, and the concatenation with another optimized measurement is the way to not waste useful information. Here, we made it through a ME measurement, from which there is no further information left (otherwise, the measurement would have not minimized the probability of error). In addition to benefiting from the features that the conventional protocol provides, namely, the error-free identification of quantum states with the maximum success probability, it is still possible to increase (sometimes, considerably) the overall probability of correctly identifying them. 

There are many other possible concatenations which depend on the specific scenario where they will be applied. The initial discrimination strategy is settled by the set of input states (are they linearly independent or linearly dependent?)\ and the required level of confidence\footnote{The confidence is measured by the probability $P(\psi_j|j)$ in associating outcome $j$ to state $|\psi_j\rangle$. Note that for UD $P(\psi_j|j)=1$ for all $j$. } to identify them. If such strategy is probabilistic, i.e., it has a nonzero probability of inconclusive results, a further measurement at the inconclusive space $\mathcal{H}_?$ (with $\dim\mathcal{H}_?>1$), will retrieve the remaining information. This measurement will be settled by the figure of merit that one wants to enhance. Here, we had linearly independent input states which we wanted discriminate without error (i.e., with confidence one). So, we implemented the optimal UD measurement. We then wanted to optimize the overall probability of correct retrodictions in this scenario. For that, we applied the ME measurement at $\mathcal{H}_?$. On the other hand, if we wanted to achieve higher confidences in the retrodictions from inconclusive results, this concatenation (UD+ME) would not be the best choice. In that case, the strategy to be applied at $\mathcal{H}_?$ would be the optimal maximum-confidence (MC) measurement\footnote{A measurement analogous to optimal UD, but designed to discriminate linearly {\it dependent} states with the maximum achievable confidence. } \cite{Jimenez11,Croke06,Weir18}. This reasoning can be extended to any set of input states as well as to any other probabilistic strategy that identifies the states with some prescribed intermediate level of confidence \cite{Hayashi08,Bagan12}. In summary, the choice of which concatenation to implement depends on the specifics of the quantum task it will be applied, but whatever this concatenation is, it will be advantageous over the conventional strategy in which inconclusive results are discarded.

As mentioned earlier, our experiment, like most optical tests of quantum state discrimination, was performed with a classical laser source. The ultimate goal was to demonstrate (the concept of) a concatenated quantum measurement in order to investigate the fundamental aspects of quantum detection theory, envisaging its future applicability in quantum tasks. For this, the transition to the quantum regime in our setup is straightforward: first, one replaces the laser source by a single photon source with a sufficient transverse coherence (e.g., using heralded single photons from parametric down conversion \cite{Neves05}), encoding the information in their transverse spatial modes, and using the polarization degree of freedom as an ancilla; then, one replaces the cameras by detector arrays with single photon counting capability (e.g., electron multiplying charge-coupled device \cite{Zhang09} or array of single-photon avalanche diodes \cite{Unternahrer16}). The stages of state preparation, coupling with the ancillary polarization, and discrimination measurements would be exactly the same as shown in Fig.~\hyperref[fig:setup]{\ref{fig:setup}(a)}. Therefore, in addition to demonstrate the discrimination protocol with concatenated measurements, our setup is potentially useful for practical applications using {\it qudits} encoded in the transverse spatial modes of single photons.

In conclusion, we demonstrated the discrimination of nonorthogonal quantum states in high dimensions by concatenating the optimal UD measurement with a ME strategy. By not discarding the information contained in the inconclusive results, we achieved significantly higher overall probabilities of correct retrodictions than in the conventional UD protocol. The concept demonstrated here can be extended to many other discrimination strategies and will enable one to explore the full potential of nonorthogonal qudit states for both fundamental and applied quantum phenomena.\\

\noindent \textbf{Acknowledgements} This work was supported by CNPq (407624/2018-0), CNPq INCT-IQ (465469/2014-0), and FAPEMIG (APQ-00240-15). M. A. S.-P. acknowledges financial support from FONDECyT (3170400). O. J. acknowledges,  financial support from Universidad Mayor (PEP I--2019020). A. D. is funded by ANID -- Fondecyt Regular 1180558 and Millennium Science Initiative Program ICN17$_-$012.

\appendix
\section{Bob's SLM and Kraus operation}  \label{appA}

Bob's SLM, sketched in Fig.~\hyperref[fig:setup]{\ref{fig:setup}(b)}, is a central element in our experiment. Ideally, it should implement the unitary $\hat{\mathcal{U}}_{\rm\textsc{slm}}$ (more specifically, the Jones matrix $\hat{J}_n$) given by Eq.~(\ref{eq:unitary}), which, together with the polarizing beam splitter (PBS), realize the Kraus operators (Eq.~(\ref{eq:Kraus})) for optimal UD. A thorough description of the calibration process for the transmissive LCD and the full characterization of the SLM operation will be provided in a separate paper. Here, we give a heuristic description of its operation based on the measured intensity and phase modulations as a function of the gray level (voltage) addressed to the device, and show the set of data that supports, with a good approximation, the SLM+PBS operation. 

\subsection{LCD calibration}
The optical modulation properties of the transmissive LCD were characterized by measuring the light intensities as a function of the gray level (\textsf{gl}) addressed to the display. The LCD is sandwiched by a polarization state generator composed of a linear polarizer and a quarter-wave plate (QWP), and polarization state analyzer consisting of a QWP and a linear polarizer. Each state of polarization is prepared and analyzed in the bases $\{\textsf{h},\textsf{v}\}$, $\{\pm 45^\circ\}$, and $\{\textsf{R},\textsf{L}\}$, totaling 36 measurements of light power with a power-meter for ${\rm\textsf{gl}}=0,\ldots,255$. We have also measured, through a double pinhole interferometer, the total phase imprinted on the incoming light by the LCD as function of \textsf{gl}. With these datasets, we are able to predict configurations for the wave plates before and after the LCD which will provide the required light modulation properties for our purposes \cite{Moreno03}.

\begin{figure*}[t]
\centerline{\includegraphics[width=1\textwidth]{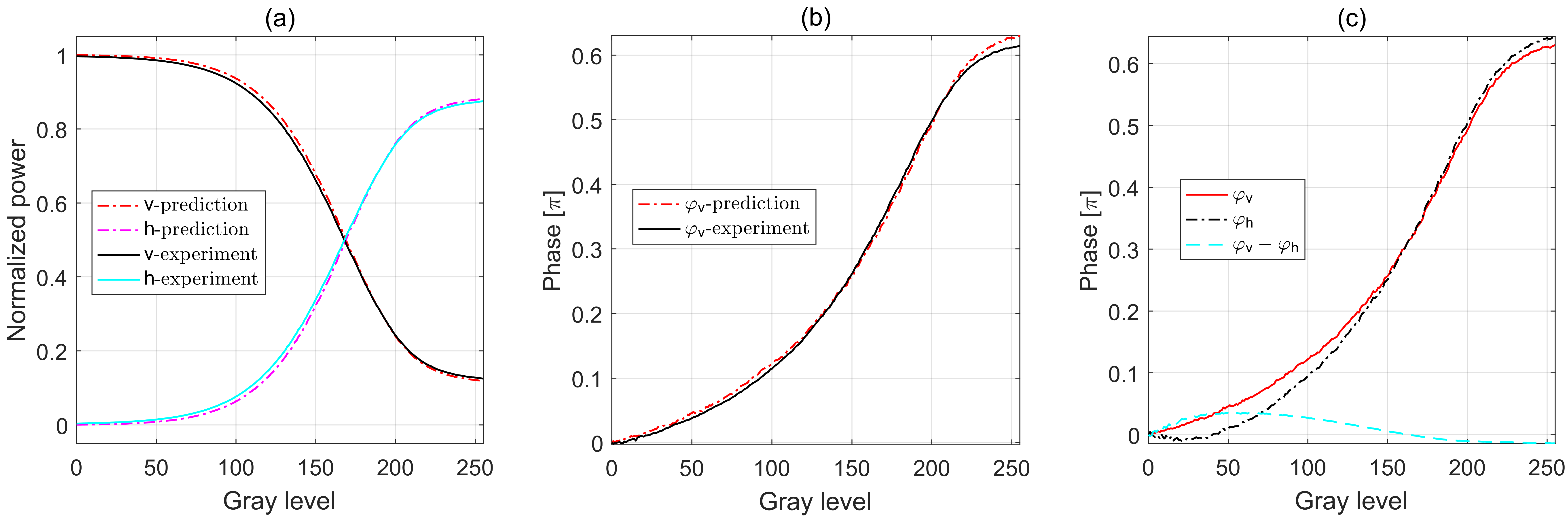}}
\caption{\label{fig:graf} (a) Normalized light power at each output of the PBS: dash-dotted lines show the predicted intensity modulations at output \textsf{v} (red) and \textsf{h} (magenta); solid lines show the experimental results at outputs \textsf{v} (black) and \textsf{h} (cyan). (b) Phase modulation at output \textsf{v} of the PBS: predicted (red dash-dotted line) and measured (black solid line). (c) Predicted phase modulations at outputs \textsf{v} (red solid line) and \textsf{h} (black dash-dotted line) of the PBS; the difference between them is shown by the cyan dashed line. }
\end{figure*}

\subsection{Requirements for the optimal Kraus operators}

The Kraus operators given by Eq.~(\ref{eq:Kraus}) act on the input symmetric states of Eq.~(\ref{eq:sym_states_in}) as follows: in the conclusive case, $\hat{A}_\perp$ removes the phases from the coefficients $\{c_k\}$ and uniformizes them according to the minimum one, i.e., $|c_k|\rightarrow c_{\rm min}\equiv\min\{|c_k|\}_{k=0}^{d-1}$ $\forall k$; in the inconclusive case, $\hat{A}_?$ also removes the phases from $\{c_k\}$ and annihilates the minimum ones, i.e., $c_{\rm min}\rightarrow 0$.

To implement this in our setup, the vertical polarization\footnote{Remind that the input state at the SLM is $|\psi_j\rangle|{\rm\textsf{v}}\rangle$.} of the light passing through the SLM must not suffer any net rotation for a given \textsf{gl} addressed to the LCD. This ensures that, after the PBS, the $c_{\rm min}$'s will be preserved at the conclusive arm and annihilated at the inconclusive one. In other words, we must find a \textsf{gl} for which the intensity of an incoming vertically polarized light will be entirely reflected by the PBS after a modulation with the SLM. In addition, between this \textsf{gl} and the one that minimizes the intensity reflected by the PBS, it is desirable to have large range of \textsf{gl}'s that produce a monotonous variation of these intensities. This enables us to implement the protocol for sets of symmetric states with varied degrees of distinguishability. Finally, the phase introduced by the SLM as a function of \textsf{gl} must be, ideally, identical at both vertical and horizontal outputs of the PBS. Denoting these phases as $\varphi_{\rm\textsf{v}}$ and $\varphi_{\rm\textsf{h}}$, respectively, it is desirable to have $\varphi_{\rm\textsf{v}}=\varphi_{\rm\textsf{h}}\equiv\varphi$, so that we can prepare the input symmetric states with the phases of the coefficients given by $\arg(c_k)=-\varphi_k$, which will be removed by the SLM afterwards\footnote{The input states of a typical discrimination protocol are known for both communicating parts.} at both outputs of the PBS.

\subsection{SLM+PBS operation}

With the LCD calibration, we searched numerically a configuration of the wave plates for which the SLM satisfy, as close as possible, the requirements described above. The best configuration we found consisted of a half-wave plate before and after the LCD (see Fig.~\ref{fig:setup}) oriented at $28^\circ$ and $29^\circ$ from the vertical axis, respectively. We measured the light power as function of \textsf{gl} at both outputs of the PBS. The normalized results are shown in Fig.~\hyperref[fig:graf]{\ref{fig:graf}(a)} where one observes an excellent agreement with the predicted modulations. At the output ``\textsf{v}'' of the PBS, the normalized power decreases monotonically in the interval $[0.1200,0.9995]$ with the increasing of \textsf{gl}. The first requirements for the optimal Kraus operators are, thereby, satisfied. The lower limit in this interval only sets the minimum possible value for $|c_k|$ in Eq.~(\ref{eq:sym_states_in}). Figure~\hyperref[fig:graf]{\ref{fig:graf}(b)} shows the measured and predicted phase at output \textsf{v} of the PBS for this configuration of wave plates. The agreement between theory and experiment is, again, excellent. The interferometric measurement performed at output \textsf{v} to obtain the phases is not suitable for the output \textsf{h} of the PBS, as the transmitted intensities are very low for \textsf{gl} up to 100 [see Fig.~\hyperref[fig:graf]{\ref{fig:graf}(a)}]. Therefore, to show that our configuration satisfy the phase requirements above, we plot in Fig.~\hyperref[fig:graf]{\ref{fig:graf}(c)} the predicted phases at each output and the difference between then. Note that in the range of \textsf{gl}'s where $\varphi_{\rm\textsf{v}}$ and $\varphi_{\rm\textsf{h}}$  show the greater discrepancies---which are still small---the intensities transmitted (or reflected) by the PBS do not vary appreciably [see Fig.~\hyperref[fig:graf]{\ref{fig:graf}(a)}]. Thus, with all these results, we can, as a good approximation, express the global action of Bob's SLM as the unitary $\hat{\mathcal{U}}_{\rm\textsc{slm}}$ given by Eq.~(\ref{eq:unitary}).

\section{Errors}  \label{appB}

\begin{figure}[h!]
\centerline{\includegraphics[width=1\columnwidth]{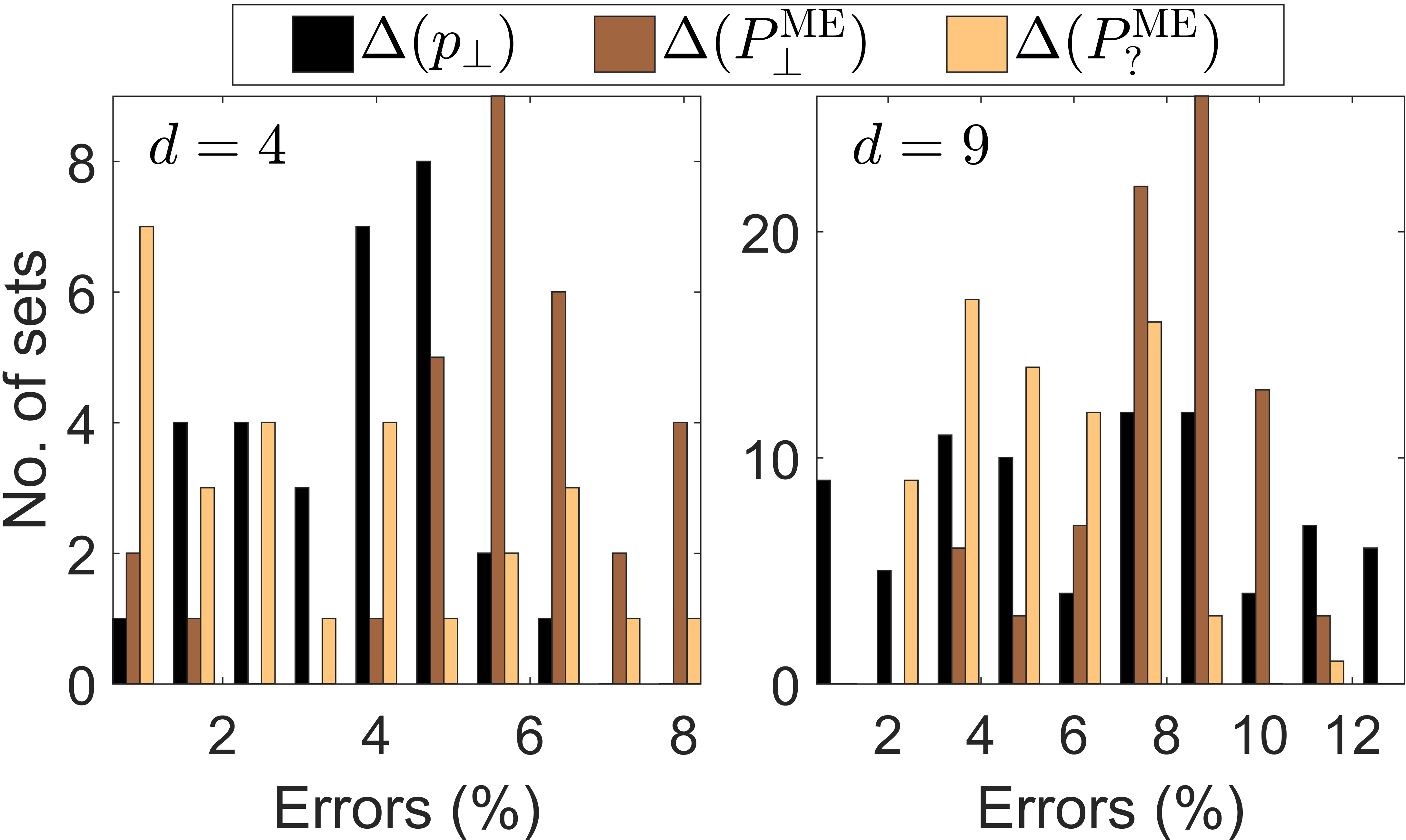}}
\caption{\label{fig:err}  Histograms of the distributions of percentage errors between theoretical and experimental probabilities for each $d$. }
\end{figure}

In Sec.~\ref{sec:Results} we presented the highest percentage errors $\Delta(p_\perp)$, $\Delta(P_\perp^{\rm\textsc{me}})$, and $\Delta(P_?^{\rm\textsc{me}})$, between the predicted and estimated probabilities of conclusive results and correct retrodictions from a conclusive and inconclusive event, respectively. We also discussed their main causes. The histograms of Fig.~\ref{fig:err} show the distributions of these errors in more details. For most of the tested sets, it can be seen that the errors got well below from the worst cases, for each $d$.  The error margins increase with the dimension, which is expected as the cumulative errors in the retrodictions grow with the number of states to be discriminated, $d$. Nevertheless, as atested by our results, these  errors have not prevented the satisfactory operation of the protocol. 

\footnotesize
\bibliographystyle{elsarticle-num}

\end{document}